\documentclass[10pt,letterpaper]{article}
\usepackage{opex3}
\usepackage{graphicx}
\usepackage{dcolumn}
\usepackage{amsmath}
\usepackage{epsfig}
\usepackage{subfigure}
\usepackage{rangecite}
\usepackage{color}

\begin{document}

%% NOTE: TITLE PAGE & TOC NOT USED FOR MANUSCRIPT SUBMISSIONS %%
%\title{Security of a discretely signaled continuous variable quantum key distribution for high rate systems}

%\vskip4pc

%\tableofcontents
%\clearpage
%% NO TITLE PAGE FOR OPEX SUBMISSIONS %%

%% START HERE
%%%%%%%%%%%%%%%%%% title page information %%%%%%%%%%%%%%%%%%
\title{A 24 km fiber-based discretely signaled continuous variable quantum key distribution system}

\author{Quyen Dinh Xuan${^1}$, Zheshen Zhang$^{1,2}$, and Paul L. Voss$^{1,2}$}

\address{1. Georgia Tech Lorraine, Georgia Tech-C.N.R.S., UMI
2958, \\2-3 rue Marconi, Metz, France \\ and \\
2. School of Electrical and Computer Engineering, \\Georgia
Institute of Technology, \\
777 Altantic Drive NW 30332-0250 Atlanta, USA}

\email{voss@ece.gatech.edu} %% email address is required

\begin{abstract} We report a continuous variable key distribution system that achieves a final secure key rate of 3.45 kb/sec over a distance of 24.2 km of optical fiber. The protocol uses discrete signaling and post-selection to improve reconciliation speed and quantifies security by means of quantum state tomography. Polarization multiplexing and a frequency translation scheme permit transmission of a continuous wave local oscillator and suppression of noise from guided acoustic wave Brillouin scattering by more than 27 dB.
\end{abstract}

 % REPLACE WITH CORRECT OCIS CODES FOR YOUR ARTICLE

%%%%%%%%%%%%%%%%%%%%%%% References %%%%%%%%%%%%%%%%%%%%%%%%%

%%%%%%%%%%%%%%%%%%%%%%%%%%  body  %%%%%%%%%%%%%%%%%%%%%%%%%%
\section{Introduction}

Quantum key distribution (QKD) systems \cite{bennett84,ekert91,nielsen00}  establish shared secret keys between two legitimate
users (Alice and Bob).  An eavesdropper (Eve) who makes
optimal physical measurements on the channel, can know, on average, none of the bits
of the secret key without inducing statistical noise that reveals her presence.  After Alice and Bob obtain samples of a correlated random
variable by means of quantum measurements, a
reconciliation phase assures agreement of Alice and Bob's data.
Finally, a privacy amplification phase uses universal hash functions
to eliminate Eve's potential partial information at the cost of
shrinking the length of Alice and Bob's shared data. Continuous variable quantum key distribution makes use of coherent quantum measurements such as balanced homodyne or balanced heterodyne detection. The result is a continuous spectrum of outcomes, as opposed to the discrete spectrum resulting from the photon counting measurements used in discrete variable quantum key distribution.
Unlike photon counters, homodyne and heterodyne detectors do not
require dead times and thus continuous variable QKD (CVQKD) systems
\cite{grosshans03,cerf01,grosshans02,lorenz04,braunstein05} are in principle scalable to
standard telecom rates, such as 10 GHz. However, coherent quantum measurements result at minimum in vacuum noise
manifesting its self as Gaussian distributed randomness with noise
power that remains constant as signals attenuate with transmission distance.  This
limits the achievable secure link length to a smaller distance than
is achievable for discrete QKD.  Thus CVQKD is best applied to distances less than 60 km.

The security of CVQKD has been studied both for Gaussian-distributed modulation\cite{grosshans05,navascues05,patron06,navascues06}, of Alice's signals and for discrete modulation\cite{namiki06,heid07,leverrier09,zhao09,zhang09}.  For a CVQKD protocol, the net secret information per channel use can be expressed as $\Delta I = \beta I_{AB} - \max(\chi_{BE})$, where $\beta$ is the efficiency of the error correcting code used in the reconciliation process. For an ideal code that achieves channel capacity, $\beta=1$. The quantity $I_{AB}$ is the classical mutual information for the channel and the signaling scheme between Alice and Bob, and $\chi_{BE}$ is the Holevo information between Bob and Eve, which bounds the upper limit of Eve's accessible information regarding Bob's measurements.  Our recent work proved the security of a CVQKD protocol designed to relax the primary bottleneck in current implementations:  the time required to accomplish reconcilation. This time is strongly influenced by the value $\beta_0$, the minimum reconciliation efficiency, which if exceeded, achieves a net positive secret key rate in terms of bits per channel use.  In order to lower the efficiency requirement and increase the speed (in bits per second) of the error correcting code that achieves reconciliation, our protocol makes use of discrete signal modulation, reverse reconciliation, post-selection, and quantum state tomography of Bob's received density matrix to bound Eve's obtainable information.

In this paper we describe an experimental implementation of this protocol.  There are currently reported continuous variable QKD experiments implemented in optical fiber\cite{lodewyck07,fossier09,qi07} and others that simulate a lossy channel with a beamsplitter\cite{lance05,symul07}. Compared to the previous work, this is the first experiment to use discrete signaling over optical fiber. This paper is also the first continuous variable system that uses a continuous-wave local oscillator (i.e. the LO and signal are not time multiplexed), which we believe to be better suited for higher speed systems. To achieve this we implement a frequency translation scheme that avoids guided acoustic wave Brillouin scattering (GAWBS)\cite{poustie92,levandovsky01}, an effect which otherwise contaminates continuous wave CVQKD. We believe this work is significant as well for potential future pulsed CVQKD systems where GAWBS noise could be present at pulse rates as low as tens of MHZ.  This paper is also the first experiment to use optical amplification in the receiver for amplification of the LO.

The paper is organized as follows: Section 2 contains a brief description
of the new CVQKD protocol, a discussion of its security can be found in Ref. \cite{zhang09}. Section 3 describes the experimental
setup and system calibration. The experimental results are in Section 4, and Section 5
contains a summary.

\section{The quantized input-quantized output CVQKD protocol}

\begin{figure}
\begin{center}
\subfigure[]{\label{Fig:encoding}
\includegraphics[scale=0.19]{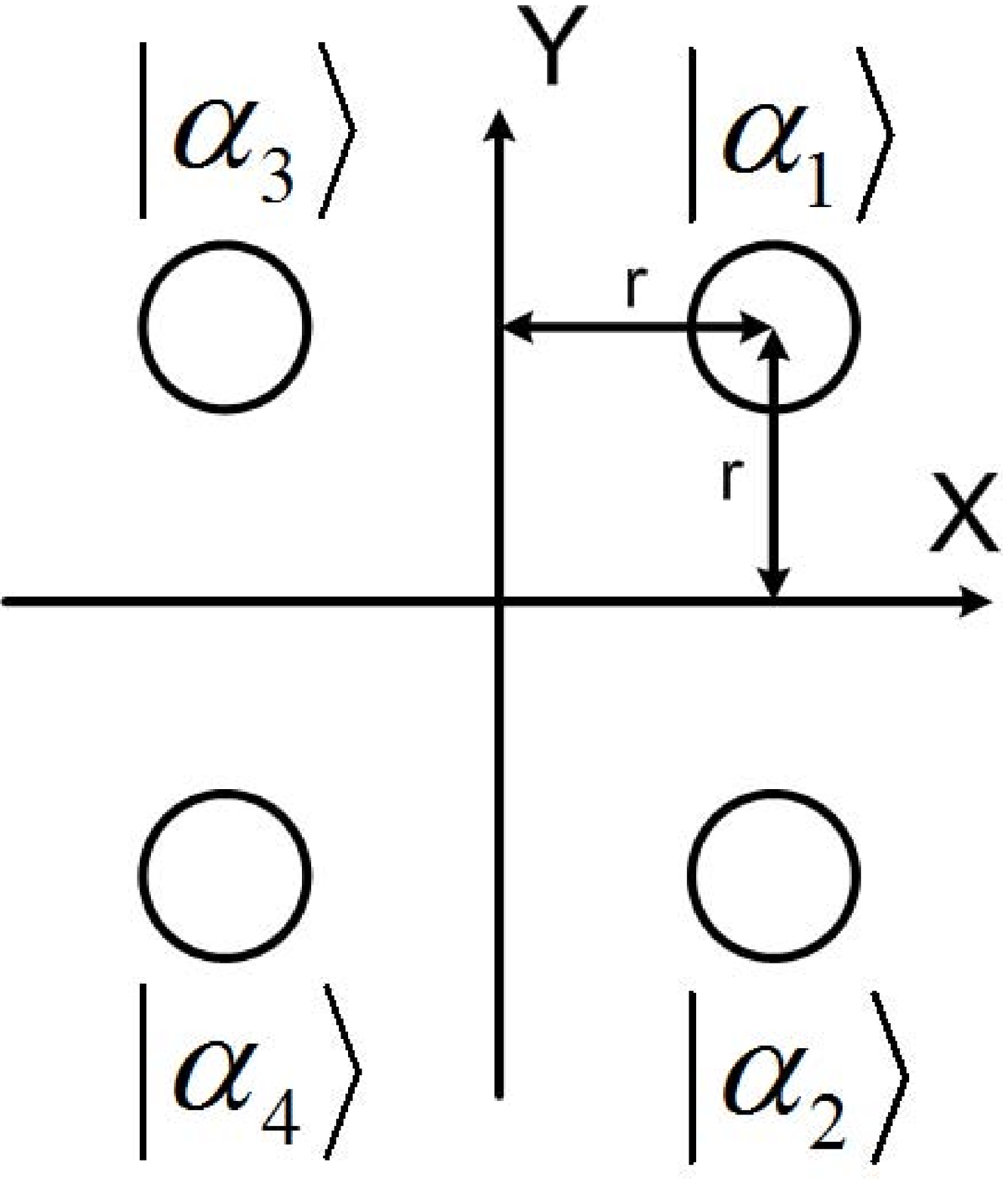}}
\subfigure[]{\label{Fig:bob_state}
\hspace{1.0mm}
\includegraphics[scale=0.92]{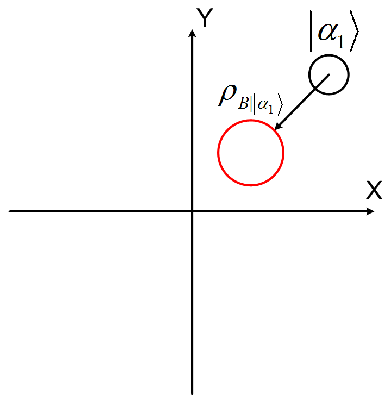}}
\end{center}
\caption{Conceptual schematic for (a) QPSK signaling, with $| \alpha_i \rangle$ chosen equiprobably and (b) security analysis, where $\rho_{B| | \,  \alpha_1 \rangle}$ denotes Bob's
received quantum state conditioned on Alice sending $| \alpha_1 \rangle$  after channel loss and possible tampering with Eve.} \label{Fig:concept}
\end{figure}

We briefly describe the protocol \cite{zhang09} in this section.  It is similar to those proposed in \cite{namiki06,heid07,leverrier09,zhao09} but  is distinguished by its use of tomographic measurement of Bob's quantum state which allows calculation of security when post-selection is used. It produces a relatively tight bound and allows for large secrecy capacity.

\textbf{Step 1:} In each time slot, Alice sends randomly and equiprobably one of four weak coherent states, as represented in Fig. 1(a), where the radius of the circle indicates somewhat arbitrarily the effective radius of quantum fluctuations.

\textbf{Step 2:} Each timeslot is randomly assigned by Bob to "data" or "tomography" subsets. For each data time slot, the measurement axis X or Y is randomly chosen by setting local oscillator phase to $0$ or to $\pi/2$ radians. For each tomography time slot, a random local oscillator phase of $0$, $\pi/4$, or $\pi/2$ is chosen. Bob also chooses a post-selection threshold.  Data having absolute value less than this threshold are discarded. Positive valued data is assigned "1", Negative data is assigned "-1".

\textbf{Step 3:} Bob now reveals which time slots contain tomography and those that contain post-selected data.  Alice then reveals which state was sent during those tomography time slots.  Using this information, Bob performs conditional quantum tomography for the four conditional density matrices, which permits calculation of an upper bound on the Holevo information for collective attacks on the system. The protocol is aborted if the the Holevo information is too large.

\textbf{Step 4:} Bob then reveals the local oscillator phase associated with each post-selected data sample. For each of these samples, Alice projects the transmitted state onto that axis.

\textbf{Step 5:} Bob sends checkbits to Alice over a public channel, i.e. reverse reconcilation.  Alice corrects her data to agree with Bob.

\textbf{Step 6:} Alice and Bob perform privacy amplification to distill the final secure key.

\begin{figure}{
\centering\includegraphics[width=110mm,height=77mm]{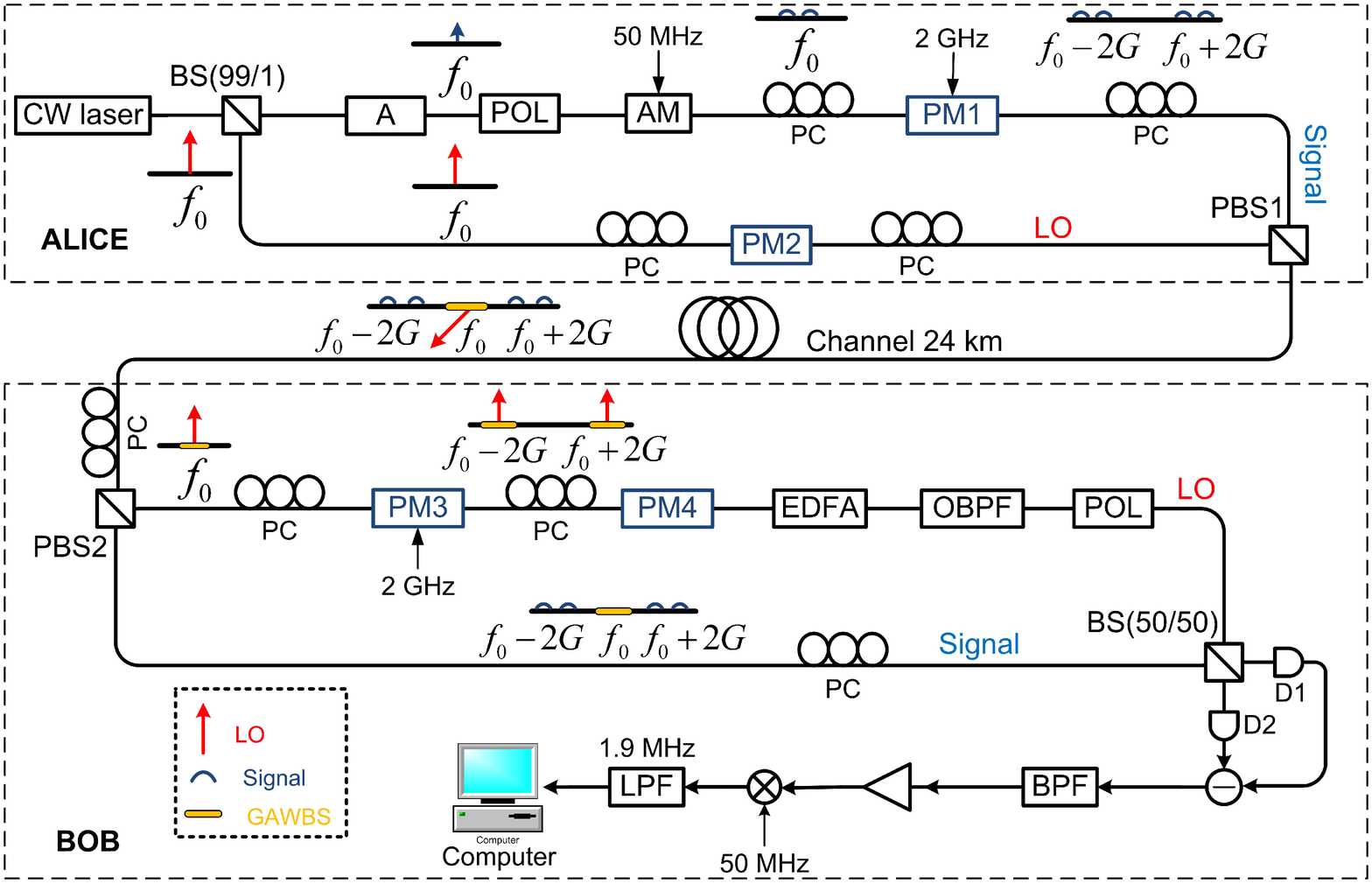}
\caption{\label{Fig:schematic} Schematic of experiment}}
\end{figure}

\section{Experimental setup and calibration process}

Alice transmits a continuous-wave local oscillator with orthogonally polarized weak coherent states, shown schematically in the upper part of Fig. 2. Phase modulator PM2 sets the phase shift between the LO and quantum signal.   Alice's information is encoded in the relative phase of the LO to coherent-state 50 MHz sidebands,  which are created with an amplitude modulator (AM) driven at 50 MHz. By biasing the AM near the extinction point the signal light is placed entirely in the two sidebands.  These 50 MHz sidebands then undergo a 2 GHz phase modulation while passing through PM1. The drive voltage for PM1 is chosen so that the phase shift amplitude $\phi$ corresponds to the first root of the Bessel function $J_0(\phi)$.  The result is that the signal light is ideally entirely frequency shifted away from the optical LO freqency creating sidebands spaced 2 GHz apart. In order to simplify Fig. 2, only two of these sidebands are represented. Two separate phase modulators are used in order to safely limit the RF power per modulator. The LO and signal are combined on a polarization beam splitter (PBS1) then sent down the transmission channel fiber having linear loss of 5.18dB.   In the fiber, guided acoustic wave Brillouin scattering (GAWBS) causes a portion of the LO to scatter into frequencies up to 1.5 GHz. These frequencies correspond to thermally populated acoustical phonon modes in the fiber that modulate the index of the fiber core. The scattering is composed of co-polarized and de-polarized components.

At the receiver, after 24.2 km of fiber, the LO power is about -6 dBm. Bob separates the LO and signal by means of PBS2, where the lengths of the LO and signal paths before recombining are matched to less than 1 mm.  Due to imperfections in the PBS, about $0.07 \%$ of the LO leaks into the signal path. The small amount of leaked LO contains detectable GAWBS noise having frequency components up to 1.5 GHz, but the frequency shifted signal remains uncontaminated as it inhabits frequencies beyond the GAWBS noise spectrum. At the LO port $99.93 \%$ of the LO enters the LO path, passing through PM3 which is modulated in the same way as PM1, ideally shifting all of the LO to sidebands. Subsequently PM4 performs Bob's random selection of phase then an optical amplifier boosts the power of the LO to 15.0 dBm, filtered by a 0.8 nm optical bandpass filter and then passed through a polarizer.  The signal and LO are mixed on a 50/50 fiber beam splitter (49.8/50.2 in practice) and guided to two photodiodes (Epitaxx ETX75) having a 1.1 GHz 3-dB bandwidth. Because the diodes have no frequency response beyond 1.4 Ghz, the different optical sidebands do not beat with each other.  The LO that specifies the single optical mode that undergoes quantum homodyne detection thus corresponds to 50 MHz double sidebands around a sequence of several optical frequencies separated by 2 GHz, but having no component at the center frequency, that of the original LO laser. A filter follows each photodiode, separating frequencies less than 5 MHz from those greater than 5 MHz.  Each photocurrent then enters a 180 degree hybrid bridge (Anzac H-9) producing an RF difference photocurrent.  The photocurrent passes through a 25 MHz highpass filter, a 50 dB-gain electrical amplifier with noise figure 0.9 dB, then a mixer that brings the 50 MHz RF frequencies to the base band. The downconverted photocurrent is finally filtered by a 1.9 MHz filter for A/D sampling. The system is controlled by a computer which uses a training frame to perform real-time adjustment of the phase drift between the signal and LO by addition of a constant to Bob's phase input to PM4. The computer also performs data post-selection and classical data correction.

The detection setup realizes between 64 and 65 dB of common mode noise suppression during experimental runs.  This balancing, reached by inducing small loss on one optical fiber and electrical path matching, is sufficient to suppress excess noise due to laser RIN noise (7 dB excess noise), GAWBS noise on the LO (15 dB excess noise), and the EDFA (10 dB excess noise). The overall quantum efficiency is 0.56, with PBS efficiency 0.795, fiber beamsplitter efficiency of 0.98, photodiode efficiency 0.74, and effective transmission losses of 0.02 due to imperfect fiber beamsplitter ratio.  During experiments, 7 mW of LO power impinges on each detector.  At this power level the electric noise is 0.069 shot noise units. Even when no quantum signal is present, there remains residual excess noise present only when Bob's signal path is connected. This noise is believed to be residual GAWBS noise and results in 0.002 to 0.005 shot noise units of excess noise remaining at the receiver. We hypothesize that the frequency translation scheme is imperfect due to uncertainty in polarization, modulation voltage, and possibly due to modulator waveguide imperfections. By comparing this to the excess noise noise present when the 2 GHz RF signal is turned off and 1.0 extra shot noise unit of GAWBS noise is measured, we estimate 27 dB of GAWBS noise suppression.  No excess noise is present when the length of the channel is 0 km and when the 2 GHz RF signal is either turned off or left on.

\begin{figure}
\begin{center}
\hspace{-0.08in}
\subfigure[Histograms for phase $\frac{1}{4}\pi$]{\label{Fig:Phase1}
\includegraphics[scale=0.125]{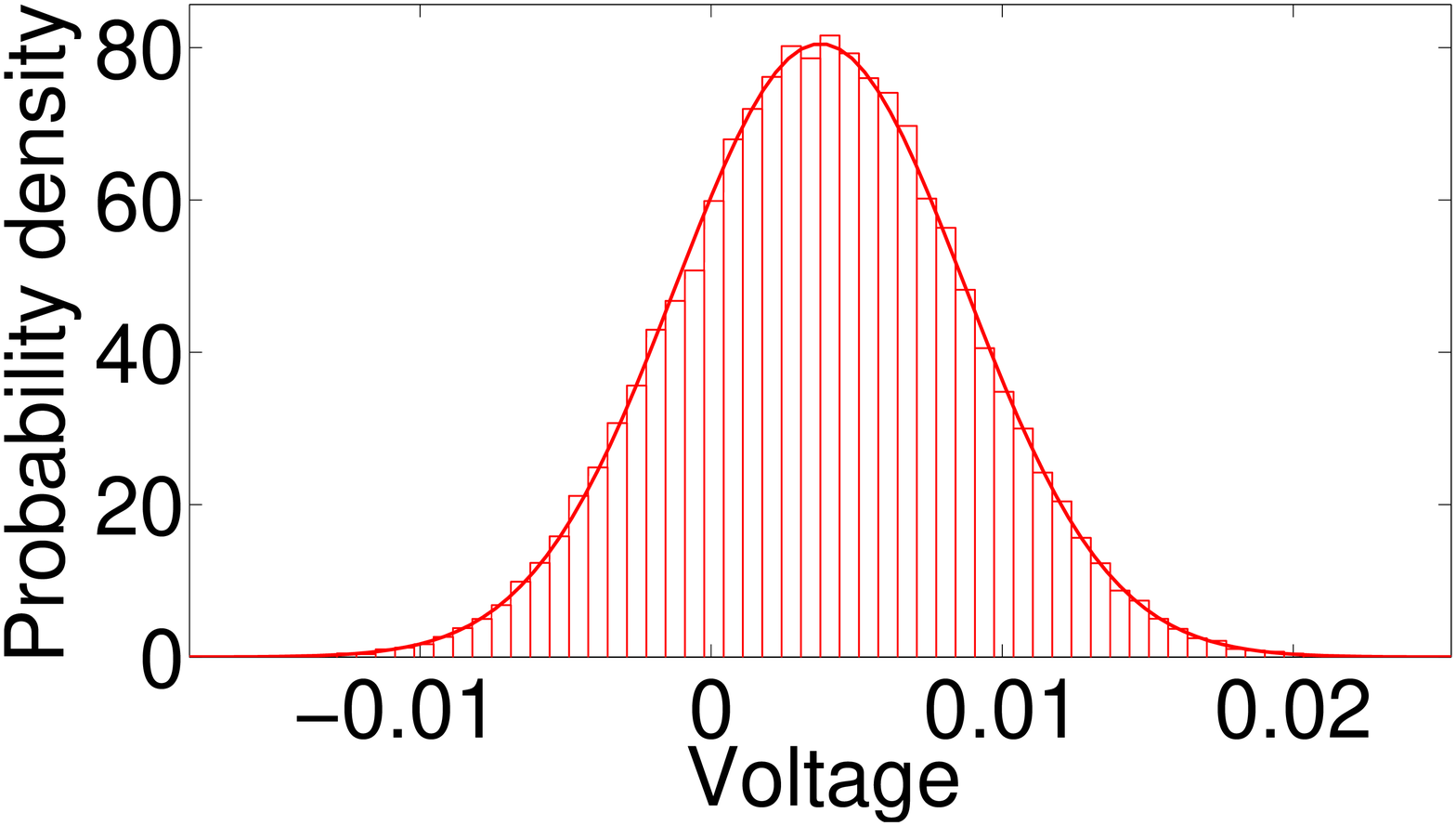}}
\subfigure[Histograms for phase $0$]{\label{Fig:Phase2}
\includegraphics[scale=0.125]{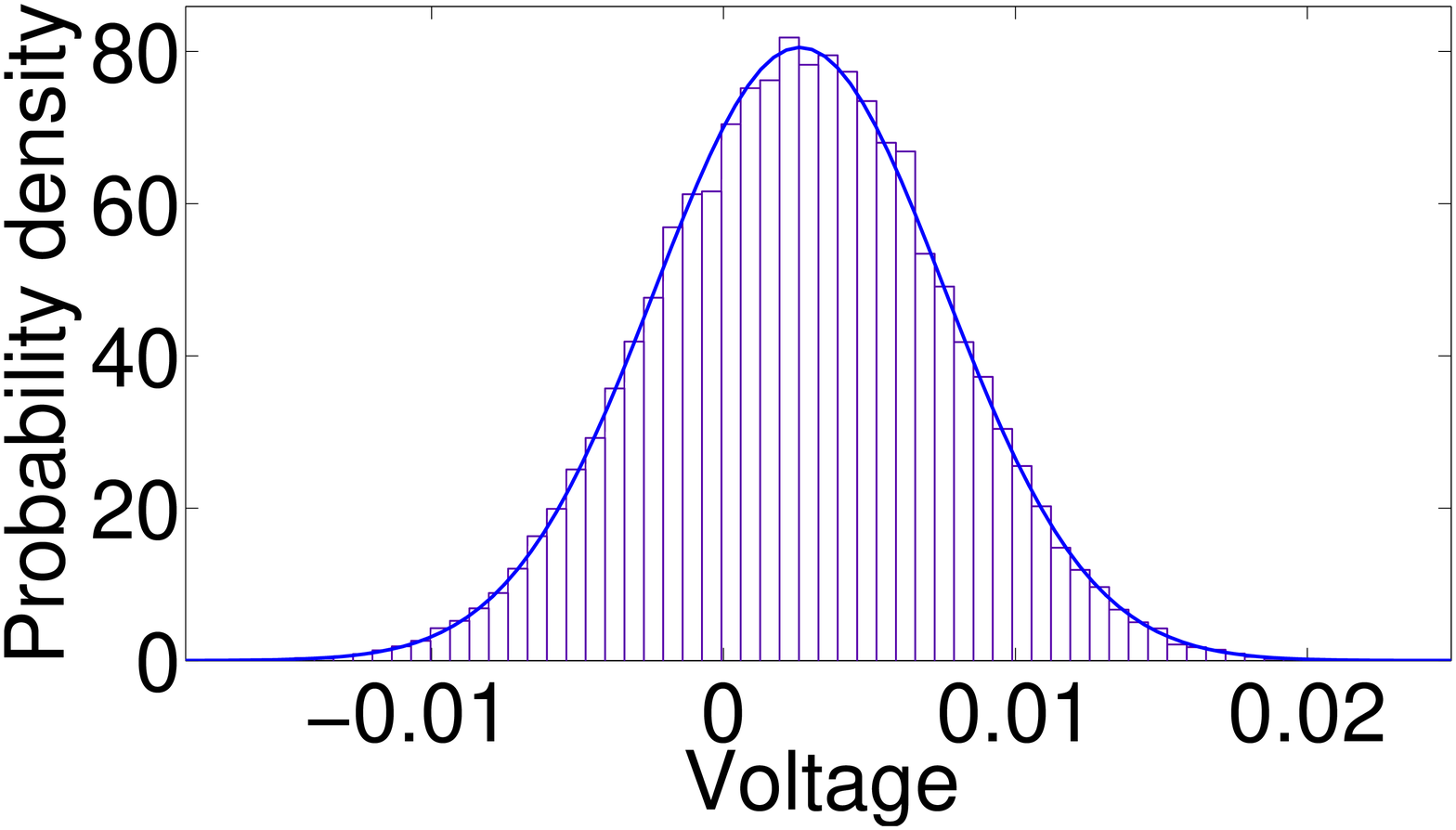}}
\subfigure[Histograms for phase $\frac{1}{2}\pi$]{\label{Fig:Phase3}
\includegraphics[scale=0.125]{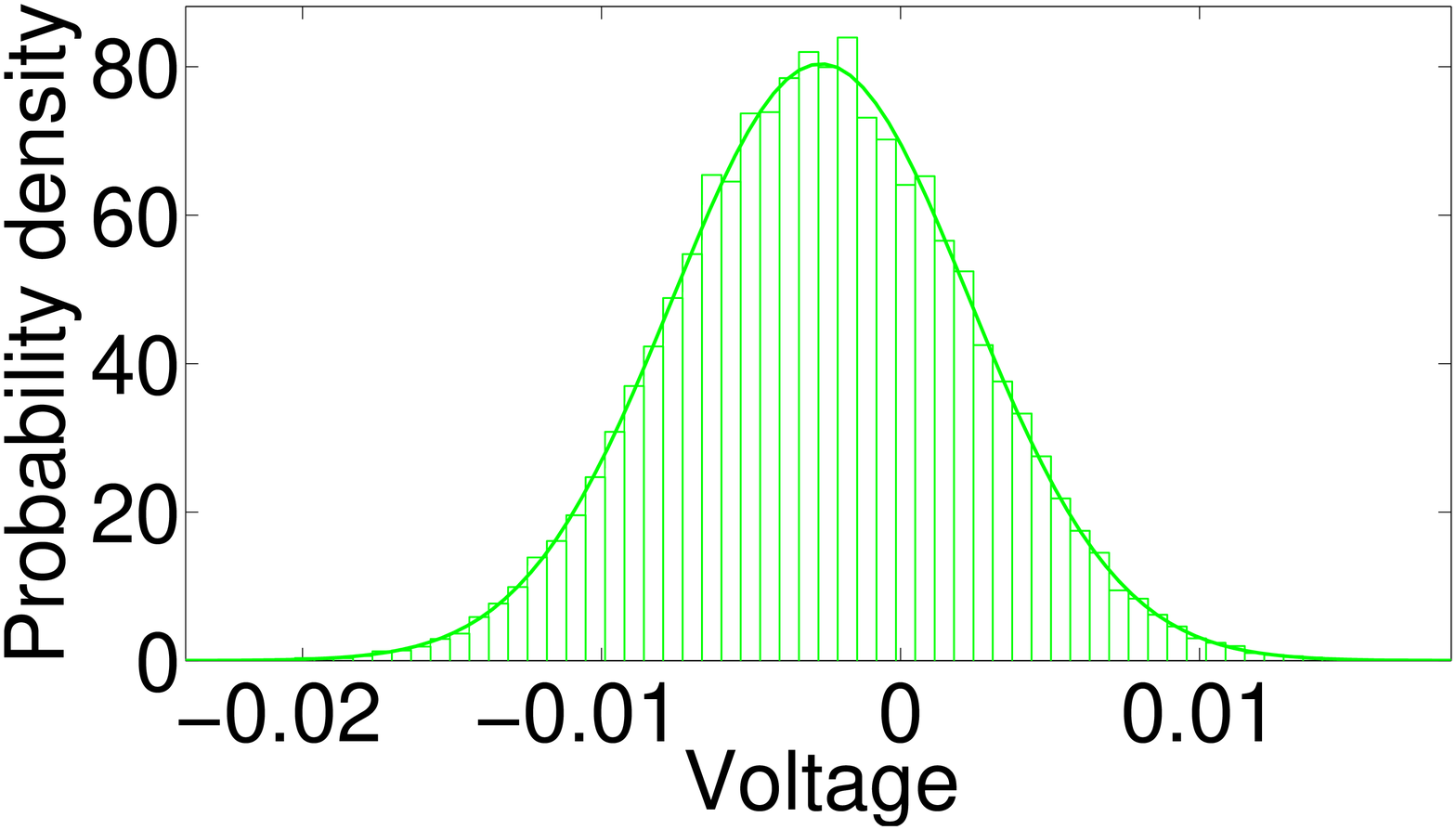}}
\end{center}
\vspace{-0.25in}
\caption{Tomography data with Gaussian fit} \label{Fig:Tomography}
\end{figure}

\section{Results and discussions}

For a $24.2$ km
channel, a received signal-to-noise ratio of  $0.272$ (ratio of signal power to standard deviation of the shot noise) and post-selection
threshold $T =1.0588$ shot noise units  meet the requirements of the error correction code used in reconciliation. Because in principle Eve could replace the communications channel with a GAWBS-free channel, adding a controlled noise-like source, the excess noise is assumed to be under the control of Eve. According to the protocol, conditional tomography for all four states has been performed.  The
results show that the average excess noise of the quantum channel at the detector is $0.0024$ shot noise units, of which $0.0024$ is due to
GAWBS noise and any remaining imperfections due to the phase estimation, amplitude modulation, and phase modulation are small and difficult to measure. In Fig.\ref{Fig:Tomography} the raw homodyne tomography histograms are shown for $10^5$ samples per phase, which show excellent agreement with the expected Gaussian distribution for coherent light with very small excess noise. For
each of the four signal states transmitted by Alice, three angles used by Bob's tomography.

Given the channel transmission,  detection quantum effiency, excess noise, the post-selection threshold, and the efficiency of the error correction code (efficiency $80\%$, error rate $7\%$), we operate in a secure region\cite{zhang09}, obtaining a final key rate of 3.45 kb/sec.  We note that unlike previous experiments, this experiment is not constrained by the time required for the error correction code, but by the data rate, which is limited by the 2 MS/s data acquisition and control card (National Instruments PCI-6115).  We have not implemented automatic polarization control at the input of Bob's PBS2, so the system operates well for 7 minutes before the polarization needs to be readjusted. The same experiment operating at a 20 MHz clock rate would leave us with a final key rate of approximately 60 kB/sec, which compares to the best current rate of 2 kB/sec for 25 km of fiber.

\section{Conclusion}
We have experimentally implemented a QKD system based on based on
discretely signaled continuous variables using a continuous wave LO that is
polarization multiplexed with the signal. GAWBS noise scattering from LO to signal was avoided by a frequency conversion technique.
A final key rate of $3.45$ kb/sec at $24.2$ km was achieved, which to the best of our knowledge, is currently the best performing CVQKD system over optical fiber.  It is anticipated that future work will include new modulation schemes to achieve higher rates and the study of finite data size effects on the security of the system.

\section{Acknowledgments}
This research was funded in part by the Agence Nationale de la
Recherche as part of the project HQNET. We thank Matthieu Bloch for
discussions regarding LDPC coding.

\end{document}